# MAGNET QUENCH 101

L. Bottura, CERN, Geneva, Switzerland

*Abstract*

This paper gives a broad summary of the physical phenomena associated with the quench of a superconducting magnet.

## INTRODUCTION

*Quench* ([1], [2], [3], [4]) is the result of a resistive transition in a superconducting magnet, leading to the appearance of voltage, a temperature increase, differential thermal expansion and electro-magnetic forces, cryogen pressure increase and expulsion. In this process the magnetic energy stored in the magnet, and the power provided by the power supply, are converted into heat in a percentage that can go from a small fraction to its totality. Superconducting magnets, operating at large magnetic fields, store large energies, and the damage potential by excess temperature is considerable. In addition, the operating current density of superconducting magnets is high (few hundreds of A/mm$^2$), the rate of joule power is large, and the rate of temperature increase is fast, so that quick action is necessary to prevent a quench from damaging the magnet. A quench must be detected rapidly, and will invariably lead to a shutdown of the power supply, and the discharge of the magnet, either by dissipation of the magnetic energy onto its own thermal mass, or externally, on a dump resistor.

The occurrence of quench, and the strategy to protect the magnet from degradation and damage, must be carefully included in the design process. A number of issues must be considered when looking into the consequences of a quench, and implementing the necessary mitigation:

- Temperature increase at the so-called hot-spot, which can degrade or permanently damage materials, and temperature gradients that induce thermal stresses and can induce structural failure;
- Voltages within the magnet, and from the magnet to ground, including the whole circuit, that could lead to excessive electrical stress and, in the worst case, to arcing;
- Forces caused by thermal and electromagnetic loads during the magnet discharge transient, where the electromagnetic load conditions may deviate from the envelope of normal operating conditions, especially in case of inductively coupled systems;
- Cryogen pressure increase caused by heating that can induce large mechanical loads on the cryogen containment, and thermally induced expulsion, to be accommodated by proper sizing of venting lines and valves.

In the next sections we will review the governing physics during the quench initiation and propagation, and apply simplifications to derive some useful scaling that relate magnet design parameters to quench indicators.

## PHYSICS OF QUENCH

The initiation and propagation of quench is governed by classical balance and circuital equations that can be written most conveniently in the form of a coupled system of partial and ordinary differential equations. Although the situation in accelerator magnets is three dimensional, we report below a version of these equations written in one dimension, along the length of the conductor. This is a most natural way to visualize the propagation of a quench, and although incomplete in terms of length and time scales, already provides a very good basis to establish simplified scaling laws. Note also that the length scales along the conductor (hundreds of m) and in the coil cross section (mm) are largely different, and a split of these scales when modeling quench, using 1-D for the direction of the developed conductor length, is quite natural.

### Equations

The temperature of the conductor $T_{co}$ is obtained from a heat diffusion equation:

$$A\overline{C}\frac{\partial T_{co}}{\partial t} - \frac{\partial}{\partial x}\left(A\overline{k}\frac{\partial T_{co}}{\partial x}\right) = A\dot{q}'''_{Joule} + p_w h(T_{he} - T_{co}) \qquad (1)$$

where we introduced averaged heat capacity, and thermal conductivity of the composite conductor, based on the area fraction $f_i$ of the component $i$ in the cross section, or:

$$\overline{C} = \sum_i f_i \rho_i c_i$$
$$\overline{k} = \sum_i f_i k_i$$

The joule heat term rises from zero when the temperature is under the current sharing temperature $T_{cs}$, to the value corresponding to the current fully in the stabilizer, above the critical temperature $T_c$:

$$\dot{q}'''_{Joule} = \overline{\eta} J^2$$

where $J$ is the cable current density and we have used an average electrical resistivity, defined as

$$\frac{1}{\overline{\eta}} = \sum_i \frac{f_i}{\eta_i}$$

In practice, the only component of low electrical conductivity in a cable is often the stabilizer, and the above definition can be simplified as follows:

$$\bar{\eta} \approx \frac{\eta_{stab}}{f_{stab}}$$

where $f_{stab}$ is the fraction of stabilizer (e.g. copper) and $\eta_{stab}$ its resistivity. The last term of Eq. (1) models the cooling, in case of the presence of a helium flow or bath at temperature $T_{he}$, through a heat coefficient $h$ at a wetted perimeter $p_w$.

For pool-boiling helium cooling, the time scale of the magnet quench is such that the temperature of the bath does not change significantly. Only at later time, as the energy is transferred to the helium, the bath can increase in temperature and pressure. In case of a forced-flow cooled cable, the behavior of the coolant during a quench can be modeled using the following simplified set of mass, momentum and energy conservation equations for the helium density $\rho_{he}$, velocity $v_{he}$ and temperature $T_{he}$:

$$\frac{\partial \rho_{he}}{\partial t} + \frac{\partial v_{he} \rho_{he}}{\partial x} = 0$$
$$\frac{\partial p_{he}}{\partial x} = -2 \frac{f_{he}}{D_{he}} \rho_{he} |v_{he}| v_{he} \quad (2)$$
$$\rho_{he} c_{he} \frac{\partial T_{he}}{\partial t} + \rho_{he} v_{he} c_{he} \frac{\partial T_{he}}{\partial x} = 2 \frac{f_{he}}{D_{he}} \rho_{he} |v_{he}| v_{he}^2 + \frac{p_w h}{A_{he}} (T_{co} - T_{he})$$

where $p_{he}$ is the pressure, and $f_{he}$ is the friction factor of the flow. Note that the above relation holds when friction dominates the momentum balance, which is usually the case in coils cooled by long pipes. Depending on the heating rate, heat transfer and flow characteristics, heating induced flow can be significant and participate to the quench propagation.

The final element is an equation for the whole electrical circuit, which consists in principle of a set of coupled coils, powered by a number of power supplies, and developing internal resistances that depend on the quench evolution. The currents in the coils **I** are most conveniently modeled solving a system of ordinary differential equations:

$$\mathbf{L}\frac{d\mathbf{I}}{dt} + \mathbf{R}\mathbf{I} = \mathbf{V} \quad (3)$$

where **L** and **R** are the matrices of inductance and resistance of the circuit, and **V** are the voltage sources provided by, e.g., the power supplies in the circuit. Capacitive effects are neglected in Eq. (3). Although the circuit capacitance can affect voltage differences, its contribution to the current waveforms is generally negligible. Note that the resistance the circuits contain non-linear resistances (the quench resistance), non-linear voltage sources (e.g. diodes), and switching actions can change the circuit topology.

The set of Eqs. (1)-(3) is strongly coupled, and in particular:

- The heat generated by Joule heat in can transfer to the coolant as from Eq. (1), which expands and is expulsed from the normal zone, as from Eq. (2);
- The flow of warm coolant described by Eq. (2) next to a superconducting wire couples heat back into Eq. (1), and is a possible mechanism of quench propagation;
- The resistance of the quenched conductor from Eq. (1) enters the resistance matrix in the circuit equation Eq. (3);
- The current in the conductor from Eq. (3) enters in the evaluation of the Joule heat in Eq. (1).

The above equations contain material properties that, as well known, are highly dependent on temperature at cryogenic conditions. In practice, an analytic treatment of the complete set of coupled equations is impossible, and one has to resort to approximations. In the following sections we will discuss such approximations.

## HOT SPOT

The main concern in case of quench is to limit the maximum temperature in the magnet. The peak temperature location, the so called hot-spot, is invariably at the location of the initial transition to the normal zone, where the Joule heating is acting for the longest time[*]. A conservative estimate of the hot-spot is obtained using the heat balance Eq. (1), by assuming adiabatic conditions, resulting in the following equation:

$$\bar{C}\frac{dT_{co}}{dt} = \bar{\eta} J^2$$

that can be integrated [1, 2, 4]:

$$\int_{T_{op}}^{T_{max}} \frac{\bar{C}}{\bar{\eta}} dT = \int_0^\infty J_{op}^2 dt \quad (4)$$

Equation (4), which is the analogous of the design method for electrical fuses, was originally proposed for superconducting cables by Maddock and James [5]. It has the advantage that the left-hand side (lhs) is a property of the materials in the cable, while the right-hand side (rhs) is only dependent on the response of the circuit.

The integral on the lhs of Eq. (4) defines a function, $\Gamma(T_{max})$

---

[*] We make here the assumption that in no other part of the inductively coupled coil system the joule heating rate exceeds the one in the portion examined. This is not necessarily the case, especially for coupled solenoids as used in MRI and NMR systems.

$$\Gamma(T_{max}) = \int_{T_{op}}^{T_{max}} \frac{\overline{C}(\theta)}{\overline{\eta}(\theta)} d\theta \approx f_{stab} \int_{T_{op}}^{T_{max}} \frac{\overline{C}(\theta)}{\eta_{stab}(\theta)} d\theta$$

that can be evaluated for the various materials used in a superconducting cable, and the approximation is valid when the composite resistivity is dominated by the stabilizer. One such evaluation example is shown in Fig. 1 for pure copper of different RRR, at zero magnetic field. The function $\Gamma(T_{max})$ can be approximated in the temperature range of interest (100 K to 300 K) by a simple power-law expression [1]:

$$\Gamma(T_{max}) \approx f_{stab} \Gamma_0 \left(\frac{T_{max}}{T_\Gamma}\right)^{1/2} \quad (5)$$

where the two constants $\Gamma_0$ and $T_\Gamma$ are fit parameters.

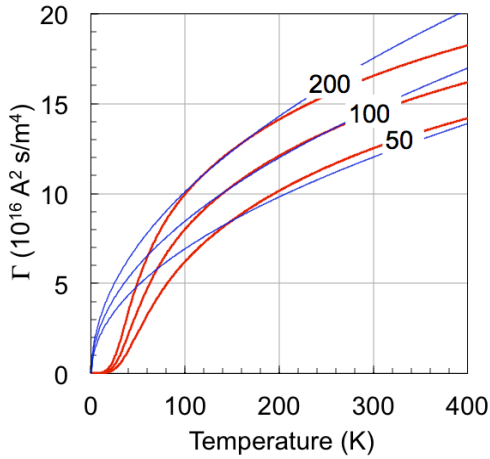

Figure 1: Sample evaluation of the function $\Gamma(T_{max})$ for copper in zero field, and taking RRR as a parameter. Also shown the power-law approximation defined in the text.

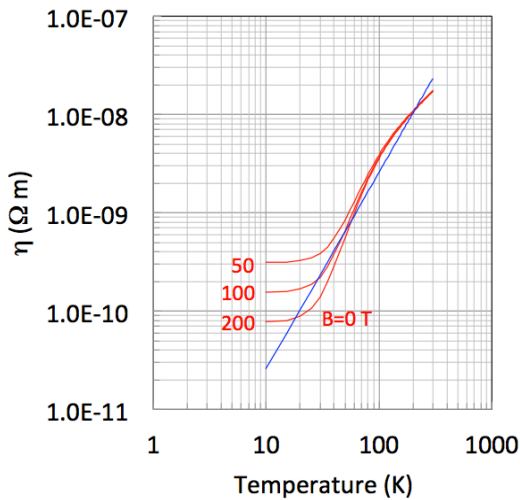

Figure 2: Resistivity $\eta(T)$ for copper in zero field, and taking RRR as a parameter. Also shown the power-law approximation defined in the text.

As we will discuss later, for the evaluation of the coil resistance during quench we also need a simple analytical expression for the stabilizer resistivity $\eta_{stab}$. This is known to be highly dependent on temperature and field. Sample data for copper are reported in Fig. 2. As also demonstrated in Fig. 2, a suitable approximation in the temperature range of interest is obtained fitting material data with the power-law:

$$\eta_{stab}(T) \approx \eta_0 \left(\frac{T}{T_\eta}\right)^n \quad (6)$$

where $\eta_0$ and $T_\eta$ are the fitting constants.

The analytical approximation Eq. (5) is much simpler to handle than the general integral, but this is not yet sufficient to allow complete analytical treatment of the adiabatic balance Eq. (4). Indeed, the rhs integral in Eq. (4) depends on the current waveform, which in the general case contains an implicit dependence on the resistivity and the size of the normal zone, i.e. the quench resistance $R_{quench}$, and on the external resistance where the magnetic energy is dumped, at least in part, i.e. the dump resistance $R_{dump}$. Suitable bounds for the current waveform can be obtained by considering two extreme cases, namely the case when the magnet is dumped on an external resistance, which is much larger than the quench resistance (*external dump*), and the case in which the whole magnet is quenched at once (e.g. by heaters) and the external resistance is negligible (*internal dump*).

*External dump*

In this case a dump resistance $R_{dump} \gg R_{quench}$ is put in series with the magnet of inductance $L$, and the current waveform is a simple exponential. The integral at the lhs of Eq. (4) yields:

$$\int_0^\infty J^2 dt = J_{op}^2 \left(\tau_{detection} + \frac{\tau_{dump}}{2}\right)$$

where $J_{op}$ is the initial cable current density, $\tau_{detection}$ is the time spent at constant current to detect the quench and trigger the system dump (including switching actions), and the time constant of the exponential dump is

$$\tau_{dump} = \frac{L}{R_{dump}} = \frac{2E_m}{V_{max} I_{op}}$$

which can be written as indicated using the magnetic energy $E_m$, the operating current $I_{op}$ and the peak discharge voltage $V_{max}$.

We can use the above approximations in the adiabatic heat balance Eq. (4) to obtain a relation for the maximum temperature:

$$T_{max} = \frac{T_\Gamma}{f_{stab}^2 \Gamma_0^2} J_{op}^4 \left(\tau_{detection} + \frac{E_m}{V_{max} I_{op}}\right)^2 \quad (7).$$

The relation above is very useful in indicating functional dependencies of the hot-spot temperature on design and operation parameters. In the case of an external dump of a magnet with given magnetic energy $E_m$ (determined by the geometrical configuration) and operating at given current density $J_{op}$ (as high as practical for winding efficiency and cost reasons) the hot-spot temperature can be reduced by:

- using materials with a large $\Gamma$ (i.e. large heat capacity, small resistivity), and large stabilizer fraction $f_{stab}$;
- detecting rapidly (small $\tau_{detection}$);
- discharging under the largest possible terminal voltage $V_{max}$;
- choosing cable designs with large operating current $I_{op}$ (decrease the magnet inductance).

Equation (7) can be studied parametrically, as shown in Fig. 3. The family of curves plotted there represent the relation between the operating current density and the maximum magnetic energy in the magnet system, resulting in a hot-spot temperature of 300 K, and taking the detection time as a parameter. The model cable parameters considered are of a Cu/Nb$_3$Sn composite with a Cu:non-Cu ratio of 1.2, operating current of 10 kA, and a discharge voltage of 1 kV. The fit parameters for the approximation of $\Gamma(T_{max})$ are $\Gamma_0 = 45 \times 10^3$ A$^2$ s/mm$^4$ and $T_\Gamma = 100$ K.

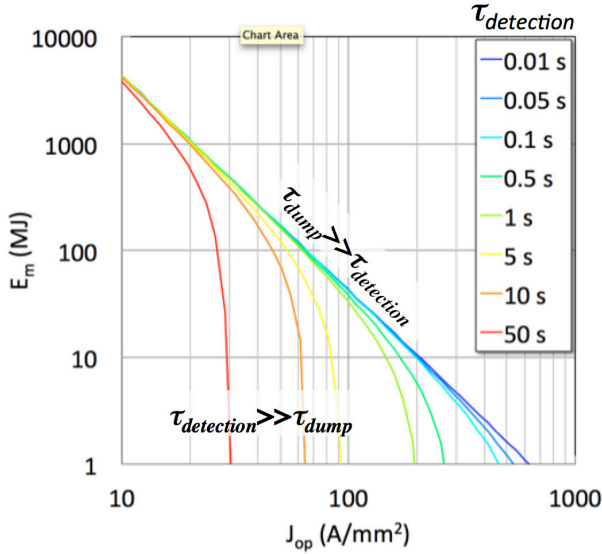

Figure 3: Case study of external dump. Relation between operating current density and maximum magnetic energy yielding a hot-spot temperature of 300 K in a Cu/Nb$_3$Sn strand with Cu:non-Cu ratio of 1.2, 10 kA operating current and 10 kV discharge voltage.

We recognize in the plot two regimes. If the detection is fast, and the energy is dissipated mostly during the dump time, the allowable magnetic energy of the system decreases like the inverse of the square of the operating current density. This is the upper envelope in the family of curves, marked in Fig. 3 as $\tau_{dump} \gg \tau_{detection}$. This limit, in practice, gives the highest possible size of a magnetic system designed for a given operating current density, assuming protection based on external dump and complete energy extraction. To fix orders of magnitude, with the parameters chosen it is not possible to limit the hot-spot temperature below 300 K in a magnet with stored energy in the range of 10 MJ and an operating current density above 200 A/mm$^2$ if the maximum discharge voltage is 1 kV. Once the hot-spot limit, the magnetic energy, and the operating current density are given, the only means to extend this limit is by increasing the operating current $I_{op}$ or the terminal voltage $V_{max}$.

The second regime is found when the dump happens rapidly with respect to the time required for detection and switching, so that most magnetic energy is dissipated by Joule heat during the latter time. This regime is the region identified in Fig. 3 as $\tau_{detection} \gg \tau_{dump}$. In this regime the hot-spot reaches the maximum allowed at the end of the detection time, under the Joule heating at constant current. This happens more or less rapidly depending on the operating current density, irrespective of the magnetic energy in the system. The limit becomes hence a simple relation between $\tau_{detection}$ and $J_{op}$, and lines in the plot become vertical. Once again, to fix orders of magnitude, the maximum allowable detection time to limit the hot-spot temperature below 300 K, at an operating current density of 200 A/mm$^2$, is of the order of 1 s, irrespective of operating current, terminal voltage, and magnet stored energy.

*Internal dump*

In the case of an internal dump, the dump resistance is negligible, and the energy is completely dissipated in the magnet system. Still, without the knowledge of the evolution of the quench resistance $R_{quench}(t)$ it is not possible to compute the current in the system, and evaluate the integral at the lhs of Eq. (4). Indeed, the general case requires the knowledge of the initiation and propagation of the normal zone, which is quite complex. To further simplify, and obtain analytical estimates, we make the assumption that the magnet is quenched completely once a normal zone is detected. This is a situation of relevance for accelerator magnets, where heaters are fired to spread the normal zone, and to hasten the dump. Following Wilson, we finally make the hypothesis that the current waveform can be approximated by a step function, with the current remaining constant for a time $\tau_{quench}$ necessary to dissipate the whole stored magnetic energy, and dropping to zero instantaneously after this time [1]. In this case the adiabatic balance Eq. (4) simplifies, as the integral of the current density becomes trivial, and using Eq. (5) we obtain the following approximation for the temperature evolution of the magnet bulk:

$$T \approx \frac{T_\Gamma}{f_{stab}^2 \Gamma_0^2} J_{op}^4 t^2 \qquad (8)$$

which holds until the time $\tau_{quench}$. To evaluate $\tau_{quench}$, we equate the joule heat dissipated in the magnet to the magnetic energy, or:

$$V_m J_{op}^2 \int_0^{\tau_{quench}} \bar{\eta}(t)\,dt = E_m$$

where $V_m$ is the volume of conductor in the magnetic system, and we remark that the only integral required is that of the conductor resistivity. At this point we make use of the power-law approximation for the resistivity, Eq. (6), and the temperature waveform given by Eq. (8) to obtain the following approximate expression for the quench time:

$$\tau_{quench} = (2n+1)^{\frac{1}{2n+1}} \left(\frac{e_m}{\alpha}\right)^{\frac{1}{2n+1}} \frac{f_{stab}}{J_{op}^2} \qquad (9)$$

where we introduced the stored energy per unit coil volume $e_m = E_m/V_m$, and the parameter $\alpha$ is a constant that depends on the cable materials and design, given by:

$$\alpha \approx \eta_0 \left(\frac{T_\Gamma}{T_\eta \Gamma_0^2}\right)^n$$

At this point, knowing the time $\tau_{quench}$, and using Eq. (8) we arrive at the estimate of the maximum temperature in the magnet bulk:

$$T_{bulk} \approx (2n+1)^{\frac{2}{2n+1}} \frac{T_\Gamma}{\Gamma_0^2} \left(\frac{e_m}{\alpha}\right)^{\frac{2}{2n+1}} \qquad (10)$$

It is interesting to note that the bulk temperature of the magnet only depends on the magnetic energy per unit volume, and material properties. The hot-spot temperature will be higher than the magnet bulk temperature given by Eq. (10) because of the time required to detect the normal zone and quench the magnet. Note that in this case the *detection* time is intended to include the heater firing and heater delay times, until the magnet is actually in normal state. Using again Eq. (8), and substituting for $\alpha$, the hot-spot temperature will be:

$$T_{max} = \frac{T_\Gamma}{f_{stab}^2 \Gamma_0^2} J_{op}^4 \left( \tau_{detection} + (2n+1)^{\frac{1}{2n+1}} \left(\frac{e_m}{\alpha}\right)^{\frac{1}{2n+1}} \frac{f_{stab}}{J_{op}^2} \right)^2 \quad (11)$$

We note in Eq. (11) two components for the hot-spot temperature, i.e. the temperature increase at constant current, which depends on the operating current density and the detection time, and the temperature increase generated by the dump of the magnetic energy in the magnet, which only depends on the cable properties and the magnetic energy per unit volume of coil (see the analogy with Eq. (10)). This second component does not depend on the cable dimensions, nor its operating current density, as one would expect from first principles.

Similarly to what done for the case of external dump, we can study the functional dependence of Eq. (11) by choosing typical magnet parameters of interest. We take the same model cable parameters, i.e. a Cu/Nb$_3$Sn composite with a Cu:non-Cu ratio of 1.2. The fit parameters for the approximation of $\Gamma(T_{max})$ are the same as before, while for the approximation of $\eta_{stab}(T)$ we take $\eta_0 = 4.1 \times 10^{-9}$ Ωm and $T_\eta = 125.6$ K. Assuming once again a hot spot temperature of 300 K, we can plot a family of curves giving the maximum stored energy per unit coil volume as a function of the operating current density, in Fig. 4.

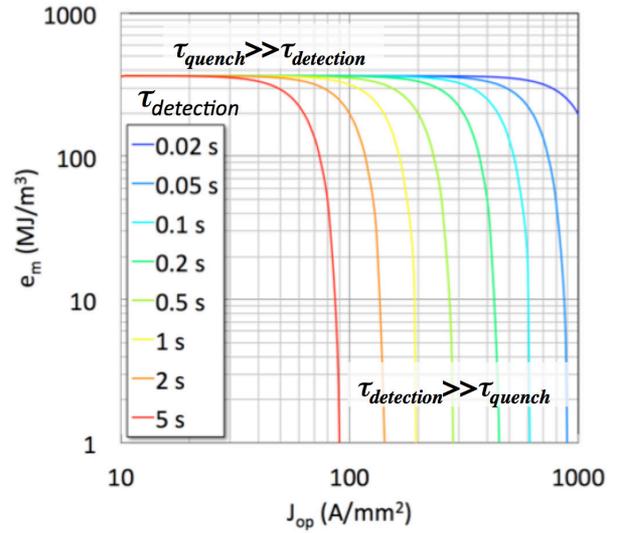

Figure 4: Case study of internal dump. Relation between operating current density and maximum magnetic energy per unit volume yielding a hot-spot temperature of 300 K in a Cu/Nb$_3$Sn strand with Cu:non-Cu ratio of 1.2.

As in the previous analysis, we can distinguish two regimes, depending on whether the magnet quench time is significantly longer or shorter than the detection time. In the first case, fast detection, identified by the asymptote marked in Fig. 4 as $\tau_{quench} \gg \tau_{detection}$, the contribution of Joule heating to the hot-spot temperature is negligible, and the limit is a horizontal line given by the energy per unit coil mass, as discussed above, independent of the cable current density. Note how the typical cable parameters chosen indicate that an energy density of 350 MJ/m$^3$ seems to be an absolute upper limit for protection, irrespective of operating current density and detection time.

The other limit is obtained when the quench time is fast with respect to the detection and heating time. This is typically the case at high current density, when resistance grows fast and the magnetic energy is dumped rapidly. In this regime, the Joule heating before detection dominates the hot-spot temperature, irrespective of the energy stored in the magnet, and the lines of constant hot-spot temperature become vertical. Note that this limit is asymptotically identical for an internal and external dump, i.e. Figs. 3 and 4 are coincident in the low energy range. As to the order of magnitude, we remark that with the cable parameters chosen, and assuming an operating current density of 400…500 A/mm$^2$, it is mandatory to detect and quench a magnet even with little stored energy per unit volume within 200 ms to limit the hot-spot to less than 300 K.

## QUENCH PROPAGATION AND DETECTION

The discussion on the hot-spot temperature scaling has shown how important it is to detect a quench rapidly, so that the heat capacity reserve can be exploited to absorb the magnetic energy stored in the system, rather than being wasted taking the external power provided from the power supply. A rapid quench trigger depends on the method used to detect a normal zone, and on the threshold setting necessary to discriminate among spurious events and a real quench. Nowadays, voltage measurements are the simplest and most direct means to detect a normal zone. It is therefore of interest to estimate the time required to see a given voltage, which gives a lower bound for the detection time defined earlier.

The resistive voltage in the initial phase of a quench grows in time because the temperature of the initial normal zone increases (which causes an increase of the resistivity per unit length), and because the quench propagates in the magnet. Making the assumption that the initial normal zone is small, as would be the case for a quench triggered by a perturbation at the scale of the Minimum Propagating Zone (MPZ) [6], we see that to estimate the detection time we need to know the quench propagation velocity.

Quench propagation has been the topic of many analytical and experimental studies. A sample of early theoretical work can be found in [7-12] and the review of Turck [13], as well as the extensive reference list of [14]. Interesting later works are the theory for super-stabilized cables [15], and the mapping of propagation regimes in force-flow cooled CICC's from [16-19]. Indeed, the quench propagation velocity depends on the conductor geometry, properties, and most important on the cooling conditions. To give a sense for the differences among the different regimes, we report below typical estimates for the quench propagation velocities calculated in an adiabatic winding, a pool-boiling winding, and a force-flow cooled winding.

The expression for the quench propagation velocity in an adiabatic conductor $v_{adiabatic}$ is the following classical solution of the conduction equation developed as early as 1960 [7] and quoted by Wilson [1]:

$$v_{adiabatic} = \frac{J_{op}}{\bar{C}} \sqrt{\frac{\bar{\eta}\bar{k}}{(T_{Joule} - T_{op})}} = \beta J_{op} \qquad (12)$$

where we used the earlier definitions for the conductor properties, and we have introduced a transition temperature $T_{Joule}$ that is generally taken between the current sharing temperature $T_{cs}$ and the critical temperature $T_c$ to account for the gradual onset of Joule heating (Wilson takes the average of the two). The above expression, which is valid only for constant material properties, has been much modified by several authors to take into account variable properties. Note, however, the interesting feature that the propagation velocity is proportional to the operating current density [20].

In the case of cooling at the conductor surface, as is the case in a pool-boiling magnet, the propagation velocity $v_{cooled}$ can be obtained correcting the above expression as follows, as detailed once again by Wilson [1]:

$$v_{cooled} = \frac{1-2y}{\sqrt{1-y}} v_{adiabatic} \qquad (13)$$

where the correction factor is given by:

$$y = \frac{p_w h (T_{Joule} - T_{op})}{A \bar{\eta} J_{op}^2} \propto \frac{1}{\alpha_{Stekly}} \qquad (14)$$

which shows explicitly the proportionality relation to the Stekly "alpha" parameter $\alpha_{Stekly}$ [21], thus recalling the fact that a quench never propagates in a cryostable conductors ($\alpha_{Stekly} < 1$).

To represent the case of a force-flow cooled conductor, we resort to the theory of quench propagation in Cable-in-Conduit Conductors (CICC's) of Shajii and Freidberg, who mapped all possible cases in a universal scaling plot [18-19]. The case of most relevance, obtained for a short initial normal zone, is that of a small pressure rise, in which case is the quench velocity $v_{CICC}$ is obtained using the following expression [18]:

$$v_{CICC} = \frac{R \rho_0 L_{INZ}}{2 p_0} \frac{\bar{\eta} J_{op}^2}{\bar{C}} \qquad (15)$$

where $\rho_0$ and $p_0$ are the initial density pressure of helium, $R$ is the gas constant in the perfect gas state equation, and $L_{INZ}$ is the length of the initial normal zone.

The expressions above have very different structure, which depends on the physical mechanism mediating the quench propagation, but they all show that at constant properties and current the quench velocity is constant in time. Any deviation from a constant velocity implies a change in properties (e.g. a quench propagating into a zone of higher or lower field), or the on-set of an additional mechanism of propagation, a quench-back.

One such mechanism is transverse propagation, i.e. a quench jump from one turn to the next, or from one layer to the neighboring one, across the coil, rather than along the conductor. Estimates for the transverse propagation velocity are complex. On one hand, the characteristic longitudinal length is typically three orders of magnitude larger than the transversal one. On the other hand, this is compensated by a similar difference in thermal diffusivity in the two directions. Orders of magnitude estimates can be obtained by dimensional analysis, resulting in the case of an adiabatic winding in the following relation [1,22]:

$$\frac{v_{transverse}}{v_{longitudinal}} \approx \sqrt{\frac{\overline{k}_{transverse}}{\overline{k}_{longitudinal}}} = \kappa \qquad (16)$$

where averages are intended over the typical unit winding cell, and we have introduced a propagation anisotropy factor $\kappa$. The reader is advised that Eq. (16) only gives a scaling, and may require large correction factors to reflect reality [22].

Given the above elements, and assuming a small initial normal zone propagating in a winding pack of sufficiently large dimension, we can estimate the voltage $V_{NZ}$ generated at constant current from a volume integral of the resistivity in the normal zone, a method devised and used extensively in [1] as well as many analytic or semi-analytic quench codes. The result of the evaluation yields the following scaling relation:

$$V_{NZ} \approx \frac{8\pi}{(2n+1)(2n+2)(2n+3)} \alpha \kappa^2 \frac{J_{op}^{4n+1} v_q^3}{A} t^{2n+3}$$

where the parameters are defined earlier. With the value of $n=2$, we obtain:

$$V_{NZ} \approx \frac{4\pi}{105} \alpha \kappa^2 \frac{J_{op}^9 v_q^3}{A} t^7 \qquad (17).$$

We can make use of Eq. (17), and one of the above expressions for the quench propagation velocity to study the dependence of the detection time for a given detection threshold. We show the results of the exercise in Fig. 5, where we have taken the same conductor parameters already used earlier, we have considered an adiabatic winding (i.e. quench velocity given by Eq. (12)) and a propagation anisotropy of $\kappa=10^{-3}$. We have reported there the detection time as a function of the operating current density in the conductor and computed using the detection threshold as a parameter in the range of 10 mV to 2 V. For reference, we have also added the evaluation of the quench velocity.

The detection time scales inversely to a power, around 2, of the current density. This is due to the combined effect of the resistivity growth with temperature and to the propagation velocity, both increasing functions of the current density. This supports the common wisdom that a quench in a high current density conductor is "easier" to detect than at low current density. At an operating current density of relevance, around $J_{op}=400$ A/mm$^2$, typical values of quench velocity $v_q=20$ m/s are obtained from Eq. (12), which is the order of magnitude observed in magnet tests. At this $J_{op}$ the resulting detection time is in the range of one to few ms, depending on the voltage threshold. This is relatively short, also because detection filters and delays are not included in this simple analysis. Indeed, as we anticipated, the scaling study does not attempt to provide exact values, but rather proper functional dependencies.

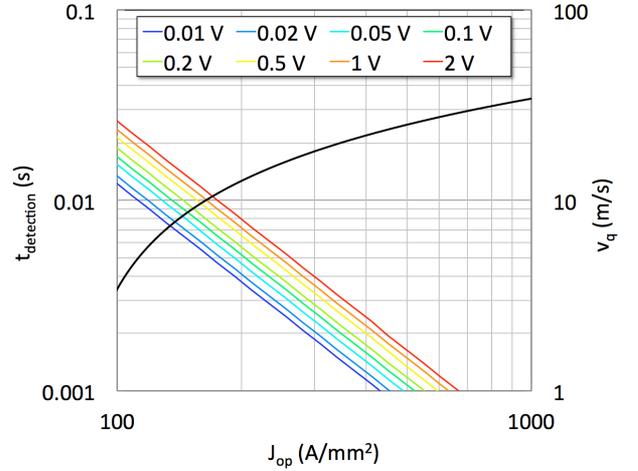

Figure 5: Relation between operating current density and detection time per unit volume yielding a hot-spot temperature of 300 K in a Cu/Nb$_3$Sn strand with Cu:non-Cu ratio of 1.2.

An interesting feature of Fig. 5 is that a variation of the detection threshold of two orders of magnitude (e.g. from 10 mV to 1 V) only results in an increase of the detection time by a factor 2 (e.g. from slightly above 1 ms to slightly above 2 ms at 400 A/mm$^2$). The reason is that once the quench is developed, the rate of temperature and voltage increase is fast (see the high power of the time function in Eq. (16)), and the difference in time among different voltage criteria is hence small. It would be interesting to test this somewhat surprising result of the scaling in a controlled experiment.

## QUENCH VOLTAGES

The resistive voltage generated in the normal zone, and the inductive voltage associated with the current variations, can be the source of a significant electrical stress in the magnetic system. An electrical failure is naturally of much concern, especially in systems of large stored energy, which is why it is important to have a good evaluation of the maximum voltage associated with a quench, both internal to the magnet (turn-to-turn and layer-to-layer) and to ground. We need to distinguish here between the two cases discussed earlier, namely the external dump and the internal dump.

In case of an external dump, the quench resistance is small with respect to the external dump resistance. The voltage seen by the coil will be maximum at the terminals and the beginning of the discharge:

$$V_{\max} = R_{dump} I_{op}.$$

The internal voltages in this case distribute in the coil according to the inductance of each portion, and can be easily deduced from the terminal voltage. Similarly, the maximum ground voltage can be obtained from the terminal voltage, once the grounding scheme of the circuit is known.

The case of an internal dump is much more complex. In this case the terminal voltage is approximately zero, as the dump resistance is much smaller than the quench resistance. The internal voltage, however, is not. The local potential results from a distributed component, the inductive voltage associated to the current variation, and a localized component, the resistive voltage in the normal zone. Unless the normal zone extends over the whole magnet (true only at late stages in the quench), and the inductance and resistance per unit length are constant in the magnet (never true), the local value of the potential can rise to relatively large values, still maintaining a value close to zero at the terminals. In this case the analysis of the voltage requires the knowledge of the extent and temperature of the normal zone. Following again Wilson, it is possible to obtain estimates by writing first the circuit equations of the whole magnet [1,3]:

$$L\frac{dI}{dt} + R_{quench} I = 0$$

which is obtained from Eq. (3), assuming a single circuit, and postulating zero voltage at the terminals. This equation is complemented by an equation for the quench voltage in the normal zone:

$$V_{quench} = R_{quench} I - M_{NZ}\frac{dI}{dt}$$

where we indicate with $M_{NZ}$ the mutual inductance between the whole magnet and the normal zone itself. $M_{NZ}$ is a function of time, according to the normal zone propagation and the geometry of the magnet, and varies from a small value, when the normal zone forms, to the magnet inductance $L$, when the normal zone extends over the whole length.

Combining the two relations above, Wilson obtains an equation for the quench voltage:

$$V_{quench}(t) = I(t) R_{quench}(t)\left(1 - \frac{M_{NZ}(t)}{L}\right) \qquad (18).$$

Without entering into the complex details of an evaluation of Eq. (18), we remark that during a quench the current $I(t)$ decreases, as well as the inductance term $(1-M_{NZ}(t)/L)$, while the resistance $R_{quench}(t)$ increases. This results in a maximum of the quench voltage during the transient, whose accurate evaluation generally requires a numerical simulation

## HELIUM PRESSURE AND EXPULSION

Under the large heating of a quench, and in case the winding is cooled directly by a bath or a flow of helium, the coolant undergoes a pressure increase which is caused either by the vaporization of the liquid, or the large decrease of density of supercritical helium as temperature increases. Considering here large-scale applications, where the amount of heating per unit coolant volume is considerable, the pressure increase can be very large, and provisions are taken to vent helium to buffers or, eventually, the atmosphere. The design and analysis of the cryogenic aspects of a quench are fairly complex matters, and are best dealt with semi-analytical or numerical simulation codes that take into account the transient energy deposition into the helium, and the process of expulsion.

It is however useful to fix order of magnitudes to give a feeling for the severity of a quench from the point of view of the cryogenic system. To do this, we give estimates for the pressure increase in a bath-cooled magnet, and an expression for the pressure increase in a quenching force-flow cooled conductor.

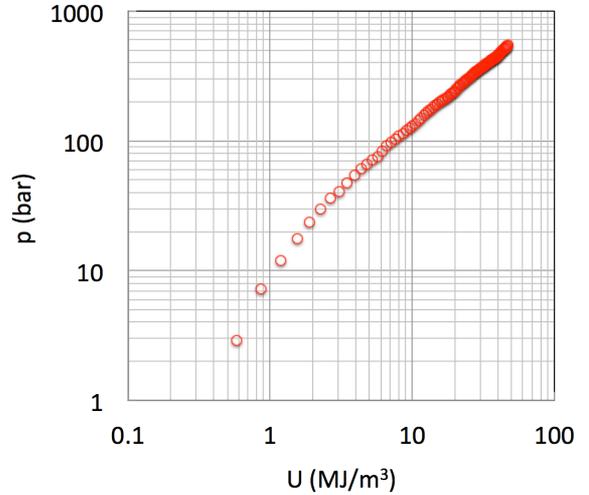

Figure 6: Relation between pressure p and internal energy density variation U at constant helium density, from an initial operating point at 1.9 K and 1 bar (case of the LHC dipole).

In the first case, we take the simple case of a bath at constant volume, which is a good approximation of the real case before the openings of quench valves. The pressure can be simply evaluated from helium thermodynamic properties, knowing the initial state, and the energy deposition in the bath. If we take the example of the LHC dipoles, with an initial state at 1.9 K and 1 bar, Fig. 6 gives the helium pressure as a function of the energy per unit volume. In the case of the LHC dipoles, the helium volume is of the order of 0.3 m$^3$, and the stored energy is 10 MJ per dipole. If all the energy were deposited in the helium bath, the pressure under constant volume condition would reach values in the range of 400 bar. Even if a small fraction of the dipole energy, less than 10 %, would be deposited in the helium, this would lead to a pressure increase above the limit of 20 bar for the cold mass, which is why a system of quench relief valves opens in case of quench, and allows helium discharge into the large buffer provided by the cryogenic lines.

In force-flow cooled magnets, the helium cannot escape freely from the quenched portion: the pressure rises and drives a heating-induced flow, which is limited by friction. Dresner developed a theory for the pressure increase in a helium pipe, and showed that the peak pressure increase $p_{max}$ in the rather conservative case of heating over the full length of the pipe is [23]:

$$p_{max} \approx 0.65 \left[ \frac{f}{D_h} \left( \frac{L}{2} \right)^3 \left( \frac{\bar{\eta} J_{op}^2}{f_{he}} \right)^2 \right]^{0.36} \quad (19)$$

where $f$ and $D_h$ are the friction factor and the hydraulic diameter of the flow, $L$ is the length of the cooling channel, and $f_{he}$ is the fraction of helium in the conductor.

Lue, Miller and Dresner [23] validated the above relation against experiments, see Fig. 7, and demonstrated that the pressure increase in a long pipe can reach very high values (hundreds of bar) under heating rates per unit volume that are applicable to the situation of a quench in a force-flow cooled conductor such as a CICC. In this case the conduit must be designed to withstand the quench expulsion pressure, or the conductor length reduced.

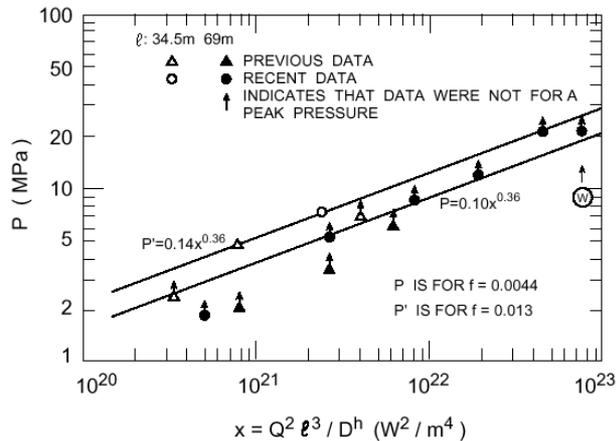

Figure 7: Experimental data on peak pressure in a heated conduit of helium, showing the results of the scaling relation Eq. (18) (from [23]).